\documentclass[prl,twocolumn]{revtex4}
\usepackage{amssymb,graphicx,afterpage}
\begin{document}
\title{\boldmath Pressure induced suppression of the singlet insulator
phase in BaVS$_3$: infrared optical study \unboldmath}
%
%
\author{I. K\'ezsm\'arki}
\author{G. Mih\'aly}
\affiliation{Electron Transport Research Group of the Hungarian
Academy of Science and Department of Physics, Budapest University
of Technology and Economics, 1111 Budapest, Hungary}
\author{R. Ga\'al}
\author{N. Bari\v si\'c}
\author{H. Berger}
\author{L. Forr\'o}
\affiliation{Institut de Physique de la mati\`{e}re complexe,
EPFL, CH-1015 Lausanne, Switzerland}
\author{C.C. Homes}
\affiliation{Department of Physics, Brookhaven National
Laboratory, Upton, New York 11973}
\author{L. Mih\'aly}
\affiliation{Department of Physics, State University of New York
at Stony Brook, Stony Brook, New York 11794-3800}
\date{version \today}
%
%
\pacs{\ }
\begin{abstract}
The metal-insulator (MI) transition in BaVS$_3$ has been studied
at ambient pressure and under hydrostatic pressure up to
$p=26\,$kbar in the frequency range of $20-3000\,$cm$^{-1}$. The
charge gap determined from the optical reflectivity is enhanced,
$\Delta_{ch}(p)/k_BT_{MI}(p)\sim 12$. This ratio is independent of
pressure indicating that the character of the transition does not
vary along the $p-T$ phase boundary. Above the critical pressure,
$p_{cr}\sim20$\,kbar, metallic spectra were recorded in the whole
temperature range, as expected from the shape of the phase
diagram. Our results exclude the opening of a pseudogap above
$T_{MI}$ at any pressure. Below $T_{MI}$ an unusually strong
temperature dependence of the charge gap was observed, resulting
in a $\Delta_{ch}(T)$ deviating strongly from the mean field-like
variation of the structural order parameter.
\end{abstract}
\maketitle

%
%
A class of symmetry breaking phase transitions, characterized by
an anomalously large gap parameter, has recently attracted much
attention and has been investigated intensely in the wider
framework of strongly correlated electron systems. Though in some
manganites and nickelates the huge enhancement of the
$\Delta/k_BT_c$ ratio, sometimes as large as $\sim30$, is
accompanied by the opening of a pseudogap above $T_c$
\cite{Katsufuji,Kim}, in most cases a mean-field-like temperature
dependence of $\Delta$ is observed. In contrast, in first order
metal-insulator transitions a more drastic, often discontinuous
opening of the charge gap is seen. On the other hand, the
transition in the inorganic spin-Peierls system CuGeO$_3$ looks
almost like a first order one, as the opening of the
singlet-triplet gap is much sharper than the BCS functional form
and follows $\Delta\propto(T_c-T)^{\beta}$ with $\beta\approx 0.1$
instead of $0.5$ \cite{Michael}. In BaVS$_3$, the detailed
temperature dependence of neither the spin gap nor the charge gap
has been measured so far. The present study of the infrared (IR)
optical properties demonstrates that electron correlations play
crucial role in BaVS$_3$: they lead to a large enhancement of the
$\Delta/k_BT_c$ ratio and simultaneously give rise to an abrupt
temperature dependence of $\Delta$.

At ambient pressure BaVS$_3$ exhibits a phase transition from a
high-temperature paramagnetic ``bad metal'' phase to a
low-temperature singlet insulator state at $T_{MI}\approx 70\,$K
\cite{Mihaly}. This is a second order phase transition, as it has
been pointed out recently by the comparison of the anomalies
observed in different thermodynamic properties \cite{Kezsmarki}.
The observation of the crystal symmetry lowering \cite{Inami} in
more recent X-ray experiments provided direct evidence for the
second order character of the transition at $T_{MI}\approx 70\,$K.

The metallic nature of the compound is enhanced by the application
of hydrostatic pressure and the transition temperature is
suppressed at an average rate of $\Delta T_{MI}/\Delta p \approx
3.4\,{\rm K}/$kbar \cite{Graf,Forro}. The critical pressure above
which the metallic phase extends over the whole temperature range
is $p_{cr}\approx 20\,$kbar \cite{Forro}. The suppression of the
insulating phase is accompanied by a monotonic decrease of the
spin gap \cite{Kezsmarki}. Moreover, the phenomenon occurring at
$T_{MI}\approx70\,$K at ambient pressure has been described as a
spin-Peierls-like transition. Its order parameter, $\Delta_{sp}$,
scales with the transition temperature as a function of pressure
according to $\Delta_{sp}/k_B T_{MI} \approx 3.6$ \cite{thesis}.
This indicates that the character of the phase transition does not
change under pressure.

Dc conductivity measurements clearly demonstrate the opening of a
charge gap ($\Delta_{ch}$) \cite{note}. However, the magnitude and
temperature dependence of $\Delta_{ch}$ can not be determined
unambiguously: the purity of the sample has a strong influence on
the transport in the insulating phase, and the activation energy
obtained from the dc conductivity is temperature dependent. The
results vary between $\Delta_{ch}=570\,$K and $1120\,$K
\cite{Nakamura,Graf}. The photoemission threshold energy obtained
by Nakamura {\it et al.} \cite{Nakamura} corresponds to an
intermediate value, $\Delta_{ch}\approx710\,$K.
\begin{figure*}[t!]
\centering
\includegraphics[width=5.5in]{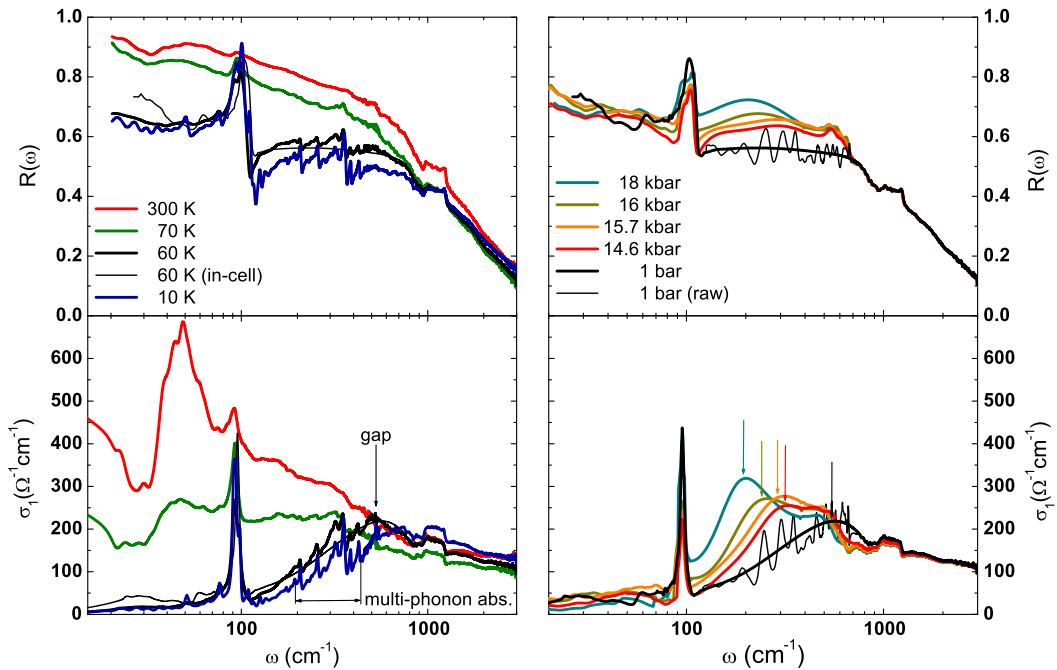}
\caption{Left: The reflectivity and conductivity spectra of
BaVS$_3$ at ambient pressure as a function of temperature. The
value of the gap and the multi-phonon branch just below the gap
are indicated by arrows. For comparison the data obtained at
$T=60$\,K inside of the pressure cell are also plotted. Right: The
reflectivity and conductivity spectra in the insulating phase of
BaVS$_3$ at several pressures. The arrows indicate the frequencies
identified with the gap values. At ambient pressure both the raw
spectra (evaluated by Eq.\ \ref{refl}) and the smoothed curves are
shown.}
\end{figure*}
The present IR optical study reveals the magnitude and temperature
dependence of the charge gap at ambient pressure.  We also find
that the $\Delta_{ch}/k_B T_{MI}$ ratio is independent of
pressure. However, in contrast to $\Delta_{sp}/k_B T_{MI}$, that
is very close to the BCS ratio, the measured $\Delta_{ch}/k_B
T_{MI}$ is $3$ times larger. The charge gap represents a higher
energy scale in the system.

Infrared reflectivity of BaVS$_3$ single crystals has been
measured in the frequency range of $\omega=20-3000\,$cm$^{-1}$,
between room temperature and $10\,$K, at several pressures up to
$p=26\,$kbar. The incident light was unpolarized, and nearly
perpendicular to the rectangular ($1\,{\rm mm} \times 3\,{\rm
mm})$, cleaved surface of the sample.  Since the small spotsize of
the beam and the high intensity in the FIR range were crucial
requirements for this study most of the measurements were done at
the U10A beamline of the National Synchrotron Light Source of the
Brookhaven National Laboratory, with a Bruker 66v/S spectrometer.
At ambient pressure the sample was mounted on the cold finger of a
helium flow cryostat. The reflectivity data were referenced to the
reflectivity of a gold mirror, also mounted on the cryostat.  Some
of the data below $50\,$cm$^{-1}$ were taken in a Bruker 113v
spectrometer, using the internal source, with a Au film evaporated
to the surface of the sample to act as a reference. The same
crystal was also studied in a custom designed self-clamping
pressure cell. Optical access was achieved through a cylindrical
natural diamond, with wedged plane surfaces to eliminate
interference fringes, and to facilitate reference measurements.
The large ($1.5\,$mm) window size and the absence of low-frequency
absorption allowed investigations down to $20\,$cm$^{-1}$.  We
used the light reflected from the outer surface of the diamond
(vacuum-diamond boundary) as a reference signal.  The sample was
mounted on the inner surface of the diamond, and the angle of
wedging between the two surfaces allowed for a clean separation of
the reference and sample reflection. Besides the optical access,
an electrical leadthrough was also implemented, and the pressure
was monitored {\it in situ} by an InSb sensor. The pressure cell
was also cooled by the He flow cryostat.

The left panel in Fig.~1 summarizes the evolution of the optical
spectra as a function of temperature at ambient pressure. The
evaluation of the spectra in the pressure cell, shown in the right
panel of Fig.~1, was done the following way \cite{thesis}. In
order to eliminate interference fringes due to the thin film of
the pressure medium formed between the sample and the diamond, at
each pressure we calculated the ratio of the insulating and
metallic phase reflectivity measured usually with $10\,$K below
and above the transition. At $p=26\,$kbar this ratio, obtained as
the quotient of the $T=7\,$K and $40\,$K spectra, is $\sim 1$ in
the whole range verifying the lack of the metal-insulator
transition. At pressures smaller than $p_{cr}$ we then evaluated
the absolute reflectivity of the insulator according to
\begin{equation}
  R_{abs}^I(p)=\frac{R^I(p)}{R^M(p)}\cdot R_{abs}^M\ ,
  \label{refl}
\end{equation}
where $R^I(p)$ and $R^M(p)$ are the raw data obtained at a given
pressure below and above $T_{MI}$, respectively and $R_{abs}^M$ is
the out-of-cell ambient pressure metallic curve, which served as a
standard. This evaluation does not influence the structure of the
spectra since, (i) in the metallic state of BaVS$_3$ the
reflectivity has a weak and monotonic temperature dependence with
a relative change $\leq 15$\% at any of the investigated
pressures, and (ii) the high-temperature dc conductivity is not
very sensitive to the pressure as $\sigma_{dc}(p=22\,{\rm kbar})/
\sigma_{dc}(p = 1\,{\rm bar}) \approx 1.2$ at room temperature
\cite{Forro}. After calculating the reflectivity by Eq.\
\ref{refl} some remains of the interference fringes are still
superimposed on the data. In order to better visualize the
results, we smoothed out this oscillation in the reflectivity by
interpolation and then applied the Kramers-Kronig transformation
as it is shown in case of the ``1 bar'', low temperature
measurement in the right panel of Fig.~1. This step effectively
lowers the frequency resolution, and it results in the smearing
out of the phonon peaks in the range of $200-400$\,cm$^{-1}$;
however, it also allows a better determination of the charge gap.
The agreement between the ambient pressure insulating spectra
obtained inside and out of the pressure cell is demonstrated in
the left panel of Fig.~1.

Above $T_{MI}$ the raw data clearly correspond to metallic
behavior: the reflectivity tends $R\rightarrow 1$ approaching zero
frequency, as clearly and directly shown in the ambient pressure
measurements.  In contrast, below $T_{MI}$ the low-frequency
reflectivity is constant. Due to the vanishing electronic
screening the phonon resonances sharpen. The dominant phonon peak
around $100\,$cm$^{-1}$ is clearly observable below $70\,$K at
ambient pressure and at every pressure as long as the metal to
insulator transition takes place. Similarly to the isostructural
BaTiS$_3$ and BaNbS$_3$ \cite{Ishii}, this peak is due to two,
closely centered, modes.  Both of them correspond to the motion of
the barium relative to the sulfur octahedron and the embedded atom
(in our case the vanadium). In terms of the optical conductivity,
the key finding is the complete suppression of the low-frequency
spectral weight below $T_{MI}$, due to the development of the
charge gap. At high frequencies the difference between the metal
and the insulator disappears and above $\sim 700\,$cm$^{-1}$ all
the curves essentially converge. At lower frequency the
conductivity goes through a gradual increase and reaches a maximum
whose frequency is identified with the gap value (pointed by
arrows in Fig.~1).  The value of the low-temperature gap at
ambient pressure agrees well with the gap derived from the
photoemission data of Ref.~\onlinecite{Nakamura}.

The single particle excitations below the gap, if exist, cannot be
distinguished from the contribution of a multi-phonon absorption
in the $200-400\,$cm$^{-1}$ range. The dominance of this phonon
branch is visible in case of the ambient pressure measurement
\cite{note2}. Note that in the same energy scale in-gap impurity
states may also give rise to an enhancement of the conductivity.

The charge gap is $\Delta_{ch} \approx 530\,$cm$^{-1}=750\,$K at
ambient pressure and it is reduced by the applied pressure as the
maximum is shifted to lower frequencies. At $p=26\,$kbar neither
the opening of the gap nor the sharpening of the $100\,$cm$^{-1}$
phonons can be detected indicating that the material remains
metallic. This is in agreement with the shape of the $p-T$ phase
boundary and the value of the critical pressure
$p_{cr}\approx20\,$kbar determined by resistivity measurements
\cite{Forro}.
\begin{figure}[th]
\centering
\includegraphics[width=2.5in]{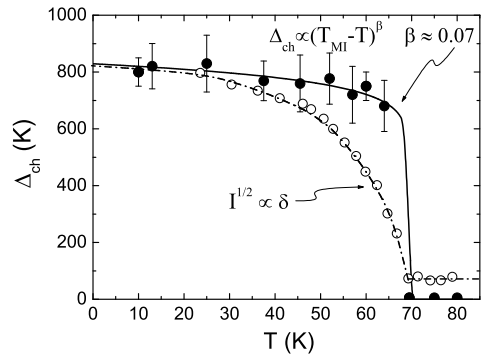}
\caption{The solid circles show how the charge gap opens at
ambient pressure and the solid curve is the power-law fit on the
data below $T_{MI}=70$\,K. The open circles are the temperature
dependence of the structural distortion measured by
Ref.~\onlinecite{Inami}. The dashed line represents the BCS
function.}
\end{figure}
As the temperature is lowered, the onset of the insulating state
is rather abrupt, and we observed only a weak change in the
reflectivity spectra of the insulator between $T_{MI}$ and
$T=7\,$K at any pressure. The temperature dependence of the charge
gap is shown in Fig.~2. The opening of the gap can be contrasted
with the temperature dependence of the structural distortion,
$\delta$ measured by Inami {\it et al.} \cite{Inami}. Below the
transition temperature the intensity of the superlattice
reflection was found to linear in $T_{MI}-T$, suggesting that the
structural order parameter has a BCS-like temperature dependence,
i.e. $\delta(T) \propto\sqrt{T_{MI}-T}$. The onset of the
insulating state much more sharp in the charge excitations; $85$\%
of the zero temperature gap is already reached at $T=0.92\cdot
T_{MI}$. In a wide range below $T_{MI}$ the temperature variation
can be described by $\Delta_{ch} \propto (T_{MI}-T)^{\beta}$ with
$\beta=0.07 \pm 0.013$.
\begin{figure}[t!]
\centering
\includegraphics[width=2.5in]{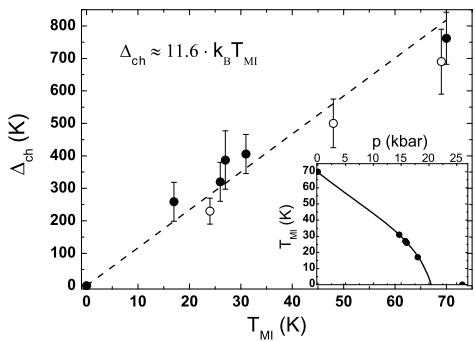}
\caption{The charge gap vs. the transition temperature as derived
from the optical (solid symbols) and the dc transport (open
symbols) experiments. The dashed line corresponds to
Eq.~\ref{ratio}. The continuous line in the inset is the $p-T$
phase boundary from Ref.~\onlinecite{Forro} and the dots show the
pressures investigated in the present study.}
\end{figure}
In contrast to the optical gap, the activation energy derived from
the dc conductivity experiments is influenced by the impurity
concentration. In a separate study we have investigated the dc
transport of the several samples under pressure. The resistivity
of the cleanest sample increased nine order of magnitude from
$T_{MI}$ down to $T=20\,$K at ambient pressure. Although its
temperature dependence slightly deviates from the Arrhenius law
one can estimate the gap within $30$\% error and find
$\Delta_{ch}=690\pm100\,$K. The magnitude and the pressure
dependence of the charge gap determined by IR spectroscopy and dc
transport agree fairly well as shown in Fig.~3. In the
$p=0-18\,$kbar range the following scaling relation holds:
\begin{equation}
  \frac{\Delta_{ch}(p)}{k_BT_{MI}(p)}\approx 11.6\ .
  \label{ratio}
\end{equation}
Finally, we shortly discuss two basic effects which could be
responsible for the large value of the gap parameter,
$\Delta_{ch}/k_BT_{MI}\sim 12$. One possibility is that the
transition temperature could be much larger, but strong
fluctuations suppress the development of the ordered state. When
the insulating state develops, the gap rapidly reaches the large
value corresponding the the "mean field" transition temperature.
In BaVS$_3$ such fluctuations may be induced by the competition of
different orderings like in La$_{1-x}$Ca$_x$MnO$_3$ ($x \approx
0.5$) where ferromagnetic and charge order (CO) coexist in a
limited region of the $T-x$ phase diagram \cite{Kim}. A wide
precursor range can also arise from the low dimensionality of the
system as it is the case in many CDW or SDW compounds or in
La$_{1.67}$Sr$_{0.33}$NiO$_4$ where the presence of fluctuating
charge stripes are observed well above $T_{CO}$ \cite{Katsufuji}.
However, the dynamical fluctuations due to the preexisting short
range order are usually manifested in the formation of a
pseudogap; this possibility may be excluded by our IR study.

Another explanation is that the charge gap is affected by the
electron-electron interaction. Such a correlation-driven
enhancement of the gap is though to be present in several
manganites like Bi$_{1-x}$Ca$_x$MnO$_3$ ($x=0.74-0.82$)
\cite{Liu}, Pr$_{0.6}$Ca${_0.4}$MnO$_{3}$ \cite{Okimoto} and
La$_{1-x}$Ca$_x$MnO$_3$ ($x\geq 0.6$) and in the nickelate
NdNiO$_3$ \cite{Katsufuji2} where the gap parameter is $\sim 20$,
$\sim 9$, $\sim 10$ and $\sim 20$, respectively. The temperature
dependence of the gap in all of the above cited $3d$ compounds
fairly follows the BCS functional form while in BaVS$_3$ the
transition appears in $\Delta_{ch}$ in a much sharper manner.  In
this sense BaVS$_3$ can rather be related to the spin-Peierls
system CuGeO$_3$, where a similarly strong temperature dependence
of the singlet-triplet gap with an exponent of $\beta\approx 0.1$
is detected by neutron scattering \cite{Michael}.  A microscopic
theory describing the metal to insulator transition and the
``pressure-magnetic field-temperature'' phase diagram
\cite{Kezsmarki,Forro,thesis} of BaVS$_3$ is highly desirable.

In conclusion, our FIR optical study supplied a detailed
experimental description of the evolution of the charge gap in
BaVS$_3$ as a function of pressure and temperature. We have shown
the lack of a pseudogap above the phase transition and a strong
deviation of $\Delta_{ch}(T)$ from the BCS-type structural order
parameter has been found. These results show that BaVS$_3$ belongs
to a novel class of correlated systems where the onset of
transition appears in a remarkable different way in the lattice
structure and the electron system.

The authors are grateful to P.~Fazekas for several indispensable
discussions. This work was supported by the Hungarian Research
Funds OTKA TS040878, T037451. A part of this work has been carried
out at the National Synchrotron Light Source at Brookhaven
National Laboratory, which is supported by the U.S. Department of
Energy, Division of Materials Sciences and Division of Chemical
Sciences, under Contract No. DE-AC02-98CH10886.

%
%


\begin{references}
\bibitem{note}In our notation $\Delta_{ch}$ is
the entire energy needed for the charge excitation, not its half.
The same convention is followed in case of the spin gap,
$\Delta_{sp}$.
\bibitem{note2}In contrast to the vibrations at $\sim100\,$cm$^{-1}$, only
the vanadium and the surrounding sulfur octahedron are involved in
these excitations. The highest energy phonon in this branch
(appearing at $350\,$cm$^{-1}$ in BaVS$_3$) has the largest
oscillatory strength. In the two isostructural materials the
spectrum are even more dominated by this mode.
\end{references}
\end{document}